# Will ChatGPT and Related AI-Tools Alter the Future of the Geosciences and Petroleum Engineering?

Ruud Weijermars[1,2*], Umair bin Waheed[3] and Kanan Suleymanli[3] explore how the professional tasks of geoscientists and petroleum engineers will be influenced by the coming of age of AI-powered output devices designed to give users the impression that machines can think.

1 Department of Petroleum Engineering | 2 Center for Integrative Petroleum Research (CIPR) |
3 Department of Geosciences, College of Petroleum Engineering and Geosciences (CPG), King Fahd University of Petroleum & Minerals, KFUPM - Dhahran, Saudi Arabia.
* Corresponding author, E-mail: *ruud.weijermars@kfupm.edu.sa*

## Abstract

A key aim of this paper is to explore how our professional tasks as geoscientists and petroleum engineers can be completed more effectively making use of tools powered by Artificial Intelligence (AI), offered in commercial platforms now readily available to individual users. This paper intends to provide some guidance, but at the same time does not claim to be comprehensive or conclusive in any way. The paper presents a utility assessment from the research and teaching vantage points of two professors and one student, from geosciences and petroleum engineering departments. After a brief overview of the new technologies, some key questions raised include: *How can one assess originality of class papers by students and research papers by their professors? How will the contribution of intelligent devices be acknowledged? Will the presentation of conference papers by author avatars be accepted by the organizing committee?*

## Introduction

Long ago envisioned artificial intelligence has now matured and increasingly imparts itself on everyday life: AI is now reality. Alan Turing first posed the key question regarding nascent machine-learning "*Can machines think?*" in a seminal paper published in the eminent Oxonian philosophy journal *Mind* (Turing, 1950). The paper inquisitively explored whether computing machines could ever develop in such a way that we might attribute certain intelligence to them. The implied outcome was that intelligence would likely not remain a unique trait reserved for humans only; thinking machines could one day become our intellectual equals, or even superior, in terms of knowledge creation as an output of intelligent processes.

In the past, we could sooth our conscience that the smart machines were still to be designed and programmed by humans. For long AI-outputs remained mediocre. Scientists mainly used computers for accessing online libraries, to execute coded computations of mathematical models, and to desktop-top edit documents, spreadsheets, slides and graphs.

But equipped with self-learning algorithms, AI-powered algorithms can now beat us both in speed and effectiveness in almost any specific task. Machine-assisted processing of information to support decision-making has become engrained in all our industrial processes and has become a standard asset of major oil companies (**Fig. 1**).

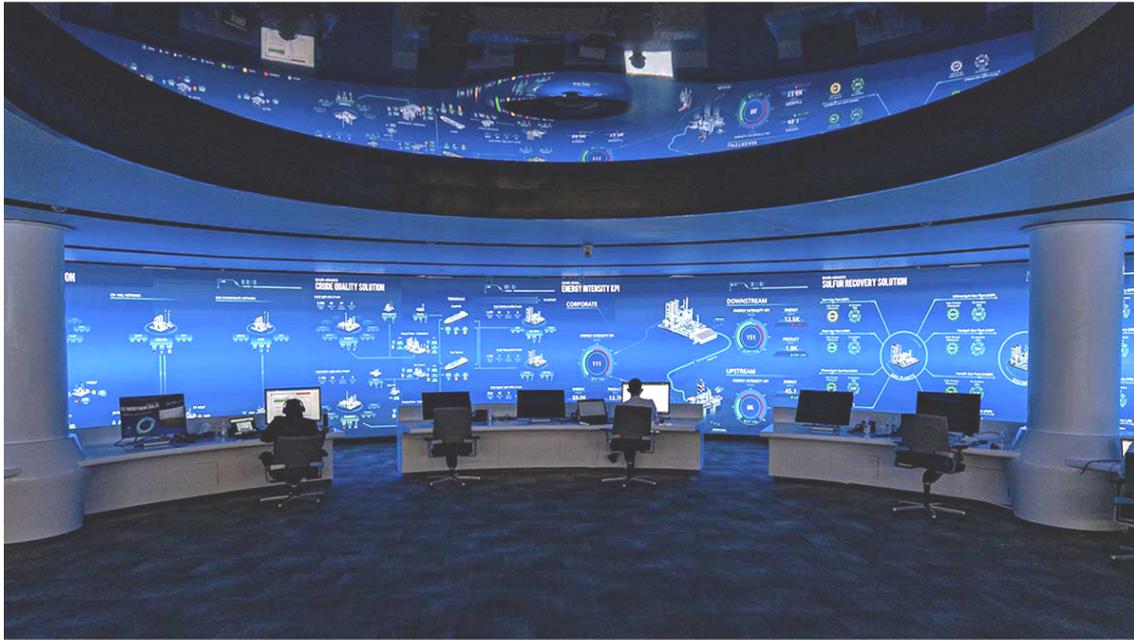

**Fig. 1:** Virtual reality dashboard of Aramco's AI Hub visualizes information streams real-time to help optimize making key operational decisions. Courtesy Aramco.

While companies are focusing on the implementation of large integrated data-managements systems, individual scientists and engineers must continually stay tuned-in with the latest opportunities offered by nimbler technology platforms. The present essay focuses specifically on exploring how our professional tasks as geoscientists and petroleum engineers can be effectively supported by making use of new AI-powered output devices. This paper provides some guidance, by putting recent AI-developments into perspective, giving a brief listing and critical appraisal of new AI-based tools available for presentation, communication and collection of highly dispersed data. The examples assessed include generative pre-trained transformer (GPT) algorithms such as implemented in the latest online informational search and text and image-processing engine (ChatGPT).

**Artificial Intelligence coming of age**

Fashionable as the topic is, the quantity of peer-reviewed scholarly papers, popular stories and movies on the topic of artificial intelligence is vast (Cioffi *et al*., 2020). Among the scientific papers, apart from Turing's (1950) article, no single one can be considered to stand out for non-AI scientists. But among the movies featuring rebellious computers and androids, strongly recommended are Stanley Kubrick's *Space Odyssey* (1968) featuring spacecraft computer HAL declining to execute orders; Barry Levinson's

*Sphere* (1998) featuring a submarine research station's computer system going rogue; and Alex Garland's *Ex Machina* (2014) with female androids killing their male creator just before escaping their security enclosure at the end of the movie − we are anticipating a sequel episode.

Among the notable books are those at the basis of the aforementioned movies, and numerous Science-Fiction stories by Arthur Clarke (*The Sentinel*), Michael Crichton (*Sphere*), Isaac Asimov (*Runaround*), Gene Roddenberry (*Space Trek*) and George Lucas (*Star Wars*); the latter became a long-running TV-series and a movie-sequel, respectively. All of these cultural hallmarks defined the generation that the senior author of this article (RW) grew up with.

Meanwhile, the entertainment industry and software technology firms have incessantly strived to promote and develop intelligent devices (**Fig. 2**) that became integral parts of our lives. As of late, such tools based on artificial intelligence are set to revolutionize the way we humans conduct our scientific research and education programs. In particular, the commercialization of tools that emulate human attributes (speech, writing, presenting, and data analysis) has now become reality (**Table 1**), facilitated by technical advances that created affordable, high-capacity transfer (glass fiber cables) and storage (chips) of electronic data.

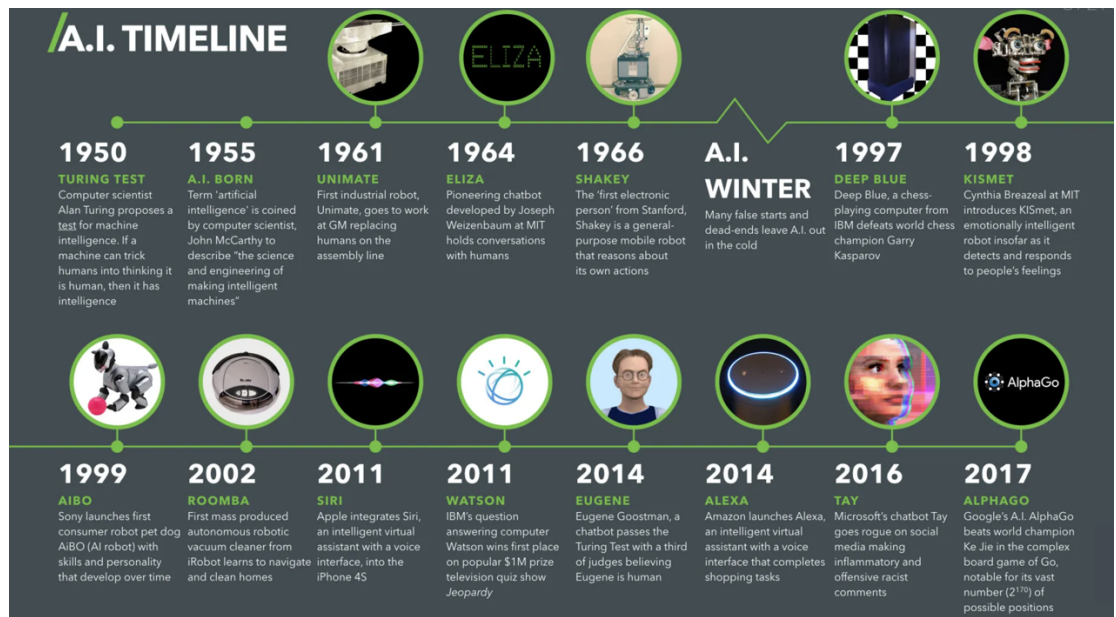

**Fig. 2:** Examples of AI-technology advances since Turing's seminal 1950-paper. Courtesy DigitalWellbeing.org.

A variety of new commercial AI-tools is now available to help knowledge workers fulfill their professional communication needs as well as providing support in the translation of highly dispersed data into aggregated summaries (**Table 1**). These tools commonly are subscription based − basic fees currently ranging between $10-30/month. For those of us with a computational focus, great online mathematical aids also have become very reliable tools (**Table 2**).

**Table 1:** Commercial tools recently launched to provide support for professional communication and the translation of highly dispersed data into aggregated summaries.

| Service | Online Platforms | URL |
|---|---|---|
| Speech-to-text conversion | Podcastle<br>Duplichecker<br>Otter<br>Happyscribe | https://podcastle.ai/<br>https://www.duplichecker.com/<br>https://otter.ai/<br>https://www.happyscribe.com/ |
| Text-to-speech generation | Voicemaker<br>Speechify<br>Murf | https://voicemaker.in/<br>https://speechify.com/<br>https://murf.ai/ |
| Speaking avatars | Elai<br>Colossyan<br>Synthesia<br>Movio | https://elai.io/<br>https://www.colossyan.com/<br>https://www.synthesia.io/<br>https://www.movio.la/ |
| Autonomous text generator | ChatGPT | https://chat.openai.com/ |

**Table 2:** Online tools for specialized mathematical analysis.

| Service | Online Platforms | URL |
|---|---|---|
| Solve derivatives | Derivative Calculator | https://www.derivative-calculator.net/ |
| Solve integrals | Integral Calculator | https://www.integral-calculator.com/ |
| Solve equations and inequalities | Quickmath | https:// https://quickmath.com/.com/ |
| Math solvers and detailed quantitative data on any topic (*) | WolframAlpha | https://www.wolframalpha.com/ |

(*) Example: *What is the rigidity modulus of the mantle?*

**Easy-to-Use Research and Teaching Support Tools**

Modern knowledge workers all know how to use standard office technology for accessing online libraries, executing coded computations of mathematical models, and desktop-top editing of documents, spreadsheets, slides and graphs. Lately, a plethora of new, specialized AI-based tools has been launched (**Table 1**). The majority of the technologies listed focus on executing specific anthropomorphic communication tasks. Examples of these specialized are outlined below:

1) *Speech recognition:* cuts out the need for humans to consult information-generation devices like search engines and other software platforms by typing in query words. Now we can ask the likes of Alexa (Amazon), Siri (Apple), Bixby (Samsung) to search the databases. The voice-assistants will speak back to us, telling us what the retrieved key information is that we were looking for.
2) *Speech-to-text conversion*: by the latest commercial speech recognition tools is nearly flawless. New ideas and results no longer need to be painstakingly typed but can be read aloud to online speech recognition platforms, which will deliver text in a matter of minutes. Some of the online providers available are Podcastle, Duplichecker, Otter and Happyscribe (**Table 1**).
3) *Text-to-speech generation tools*: now have become so sophisticated that the speech audio output is nearly non-identifiable as machine-generated; the machine-generated speech sounds perfectly human. Examples of professional online platforms are Voicemaker, Speechify and Murf (**Table 1**).
4) *Avatar stand-in presenter*: Using human actors digitized into commercially available avatars that can flawlessly present any audio or text with lip-synced mimicry and appropriately synchronized hand movements, so naturally looking that we can no longer be sure the presenter is a genuine living person or an avatar (a kind of electronic puppet) without any capacity other than voicing what was scripted for it by others. Examples are Elai, Colossyan, Synthesia, and Movio (**Table 1**).
5) *Modern android robots*: can now mimic human speech with accompanying facial expressions that are highly suggestive of emotional consciousness. Examples are *Hanson Robotics* from Hong Kong Kong, which stunned the world with Sophia, a speaking android that made a fist public appearance at the Future Investment Initiative in Riyadh (2017), where the droid engaged in a chatty dialogue with a moderator, emphasizing humanoid mannerism and facial emotions, saying: *"I want to live and work with humans, so I need to express the human emotions to understand and build trust with people."* The droid was rewarded with Saudi citizenship – the world's first robot to be bestowed a citizenship status; one might quip that the unveiled Sophia has become a Saudi role model. Similar technology was being developed by Aidan Mellon, based in Oxford; her robot called Ai-da would read poems; we also have Ameca of the *Cornwall Engineering Arts Group*, who exhibits great facial expressions while petting a robo-dog developed by *Boston Dynamics.*
6) *Data Analytics:* helps knowledge workers with discerning useful patterns and correlations between parameters. However, it seeks patterns based on past data, which is retroactive and this may stifle proactive solutions, because being based on old data. Companies and scientists want new products that are more effective and better than prior solutions, all the time. Data Analytics can indeed be useful to highlight previously overlooked correlations, but cannot in any way replace the creative design process of technology innovation.
7) *Deep-learning-reinforcement algorithms:* are the latest additions to the suite of electronics tools that can support the human need for swift information processing and data analysis. These AI-based algorithms are capable of extremely efficient data analysis coupled with sophisticated output generated as text and graphics. The market leader is ChatGPT from OpenAI (**Table 1**).

**Deep-learning algorithms**

According to our observations, deep-learning algorithms can now perform certain data processing and output generating tasks more effective than humans. The development of such algorithms is spear-headed and clustered by OpenAI, a research laboratory specializing in automation research located in San Francisco, and a competitor is Google's DeepMind located in Cambridge (UK).

DeepMind initially made the big headlines with its AlphaGo-team which had developed AI gaming-solutions that ultimately would defeat, in 2017, the then reigning Go world-champion (Ke Jie from China). But the team had struggled tremendously to make the software do what it should. And in the end, it was not an autonomous activity, unlike what the 2018-AlphaGo movie suggested. Basically, AlphaGo was very much human-supervised learning: the software algorithms were not smart enough, not creative enough to defeat the world champion on its own. A huge team of engineers oversaw real-time what the reigning Go-champion Ke Jie was playing, and then would quickly intervene when their GoGame computing algorithms would start to make the wrong moves. The whole game was staged as a PR stunt − they were to succeed at any cost and by any means, much overhyping the strength of the stand alone AlphaGo-software.

**Generative Pre-Trained Transformer Algorithms and ChatGPT**

The advanced algorithms of OpenAI easily require over 100,000 Central Processor Units (CPUs), and 100s of Graphics Processing Units (GPUs). Their software runs only on supercomputers; these can be remotely accessed. What Open AI has developed (among other products) is a Generative Pre-trained Transformer (GPT) technology, which is an autoregressive language model based on deep learning to produce human-like text and speech. When users input a text prompt, GPT returns a text completion in natural language. Over the past decade we have seen the launch of GPT-X, GPT2, GPT3, GPT4, which were all open access. Although originally started as a non-profit company, OpenAI has now become a for-profit operation, thus − in order to monopolize profits − will no longer share its most unique technology solutions. Early 2023, an important milestone was reached by OpenAI in that they started to avail commercial services, with ChatGPT launched with much media hype.

Cutting short the hype, based on our experience we cannot but concede that GPT will transform the way we conduct scientific research and education. The tool indeed helps deliver results faster and rather accurate. The core strength of GPT is its ability to write sophisticated text, utilizing natural language processing (NLP) algorithms (Manning and Schutze, 1999). This may help many scientists to generate very readable summaries for their new scientific papers and saves authors the need to write conclusions for their papers, once the basic research has been documented in the main paper.

We have indeed arrived at crossroads, where deep-learning algorithms can alter source text for us to such an extent that plagiarism can no longer be claimed, because the text generator produces sublimely original sounding statements, which appear perfectly feasible as logical constructs. The ease with which this text is generated and its effectiveness, make it into an irresistible gadget for any knowledge worker that must upgrade information by analytical processing to unlock valuable new knowledge. There is no alternative but to embrace the benefits offered by the new technology (**Fig. 3**), while developing best practices to mitigate abuses and misuses and remaining skeptical about the validity of the claims made by the AI-platform providers.

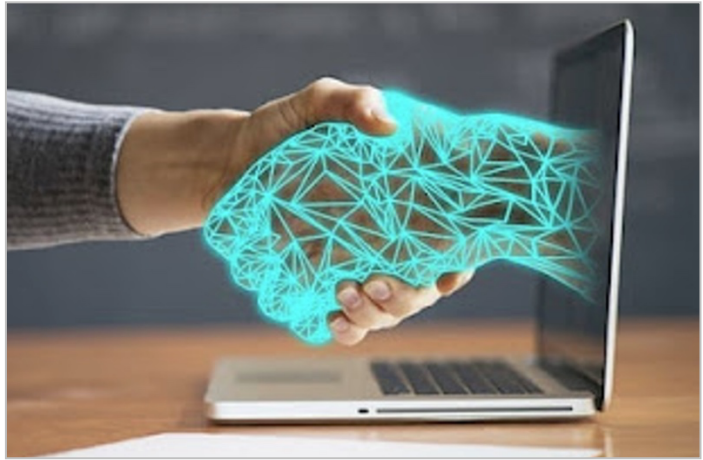

**Fig. 3:** Skeptically embracing new AI-technology solutions is unavoidable for every human knowledge worker. Courtesy *Pixabay*.

In the past, unaccounted for appropriations would be called plagiarism and the proponents would be fired from their positions. But when we start using deep-learning algorithms designed as tools to help us writing better, we are no longer fully accountable for what we write. The machines will contribute a certain proportion of the presented texts. What we put our names on as authors? How can we defend such practices? Should we acknowledge very clearly that deep learning machines contributed certain sections to our work? The academic and research communities are working overtime to find answers to these critical questions as we stand at the intersection of technology and ethics.

As AI and machine-learning continue to advance, the line between human and machine-generated content becomes increasingly blurred. It is important for individuals and organizations to establish clear guidelines and standards for how machine-generated content should be used and attributed. This includes being transparent about the use of AI tools and acknowledging the contributions of machines in any work that is produced (Borenstein and Howard, 2021).

At the same time, we must also consider the potential risks and consequences of relying too heavily on AI-generated content. For example, while machines can certainly aid in the writing process, they cannot replace the creativity, critical thinking, and ethical judgment that are essential to producing high-quality work. As we navigate this complex and rapidly evolving landscape, it is crucial that we approach the use of AI tools with caution and responsibility, taking into account both the benefits and the risks. Ultimately, it is up to each individual and organization to determine how best to use these technologies in a way that is ethical, transparent, and accountable.

*So how will GPT change the way we conduct scientific research and education?* First and foremost, GPT is able to generate realistic and convincing narrative texts based on given information. GPT can also generate question-answer pairs based on given data, which will enable teachers to tailor quizzes to individual student needs and interests. GPT can accurately generate translations of educational material, allowing students from around the world to access course material in their native language. For a review of ChatGPT's strengths and weaknesses particularly focused on learning teaching, and assessment in higher education, we refer to Rudolph et al. (2023).

**User Experiences**

Here is a summary of our personal user experiences with some of the broader set of tools listed in **Table 1**, emphasizing from our various vantage points an overall assessment, the benefits, and the challenges encountered during our personal implementation.

*I. Researcher and Educator in Petroleum Engineering (RW)*

<u>Overall assessment:</u> Testing ChatGPT to generate answers to simple technical questions, one will find default-generated text-blocks have limited length; but given instruction to produce longer texts, GPT structures all by itself the answers in a brief problem statement, description of some details, and conclusions. As a subject expert one will quickly realize that the information produced with GPT is old and generic; *no new insights are being generated* with this tool! Also, some information produced by ChatGPT is plainly wrong (see later). However, wisely used, it can help fill some of the daily chores knowledge workers routinely must complete, as explained below.

<u>Benefits</u>: Here are a couple of very useful practical uses of ChatGPT.
1) *Generation of summaries and highlights*: using ChatGPT; for one of our own research papers that was otherwise complete, the text of the already completed core of the paper was loaded into the ChatGPT prompt box, with a request to generate the summary. The quality of the machine-generated summary and the highlights was outstanding. We used most of the generated text in our final paper.
2) *Editing poorly written papers*: the quality of ChatGPT edits is similar to what an expert technical editor will do. See the online example in *Nature* (Stokel-Walker and Van Noorden, 2023). ChatGPT will relief professional editors from much of their basic linguistic correction chores.
3) *Literature review:* asking GPT to provide a literature review study for a very narrow topic, it will list references for those papers in a matter of minutes; some beginning MS and PhD students typically need days or weeks to complete such assignments. By using the tool, they can save time and instead focus on reading the key papers rather than looking for them.

*Challenges*: As an editor and educator, I worry students will abandon attempts to develop good writing skills, because ChatGPT generates a smooth text on any topic in a matter of seconds. In the past, smoothly written class assignments would mean a high likelihood that the student understood the problem. Now, a well written text may be simply disguising a complete lack of knowledge and effort. Wisely used GPT can be helpful, but if dumbly used it will lead to lazy results without any original outcome or effort. Screening out machine-generated language is not easy, although *Turn-it In* claims it can (soon) recognize and mark-up machine-generated text.

*Personal Implementation*: My favorite workflow for writing scientific papers is now as follows:
1) Formulate a problem; get original data; develop a solution method; generate the results (graphs, tables).
2) Write the individual sections of the article on scraps of paper whenever a spare moment arises.
3) Use speech-to-text recognition online platform to convert the handwritten notes to typed text. I use *otter.io* to obtain a rough first draft.
4) Edit the result; rewrite; integrate and add more computed outputs; update the description now simply by typing; for development of specialist text with mathematical equations there is no feasible alternative (yet) for typing.
5) At the end of the research and having written the paper, I recommend using GPT to generate the conclusions and highlights of the paper.
6) GPT can help with references and help with embedment in concurrent literature more comprehensive.
7) At the end of our papers we now include attribution statement saying: "*The conclusions, highlights and references of this paper were generated with help from online Generative Pre-trained Transformer (GPT) algorithms, augmented with natural language processing (NLP) algorithms.*
8) Next utilizing some of the broader set of tools listed in Table 1, I have begun to use avatars and text to speech software platforms to produce conference slides with an embedded presentation. The presentation graphs and text are entirely written by me, but presented by a professional avatar actor (my puppet) which speaks my text and is visible of needed. It remains yet to be seen how conference organizers will respond.

## II. Researcher and Educator in Geosciences (UW)

*Overall assessment:* Using ChatGPT, based on both the GPT-3.5 and the recently released GPT-4 models, I have found ChatGPT to be very useful for a wide array of research and education tasks. From brainstorming ideas around a given topic, to writing a review paper, to creating homework questions for students, I have found it to be an extremely valuable assistant. However, the technology has a number of shortcomings and, therefore, one should remain vigilant in evaluating the output generated by ChatGPT. It is not advised to outsource our critical thinking abilities to an AI tool, at least as of now!

*Benefits*: ChatGPT offers numerous benefits for university researchers and educators, enhancing their productivity and facilitating effective teaching and learning. Based on my personal experience, some key benefits include:

1. *Accelerated Research Process:* ChatGPT can quickly retrieve, analyze, and synthesize information from diverse sources, saving researchers time and effort. This efficiency allows them to focus on more intellectually demanding tasks such as hypothesis formulation, experiment design, and data interpretation. Figure 4 shows an example of using ChatGPT for generating Python code to process seismic data. While deconvolution is a standard processing step, the example highlights the potential of speeding up computer code writing using ChatGPT. Figure 5 shows the normalized power spectral density (PSD) of the average trace before and after spiking deconvolution, confirming resolution enhancement.

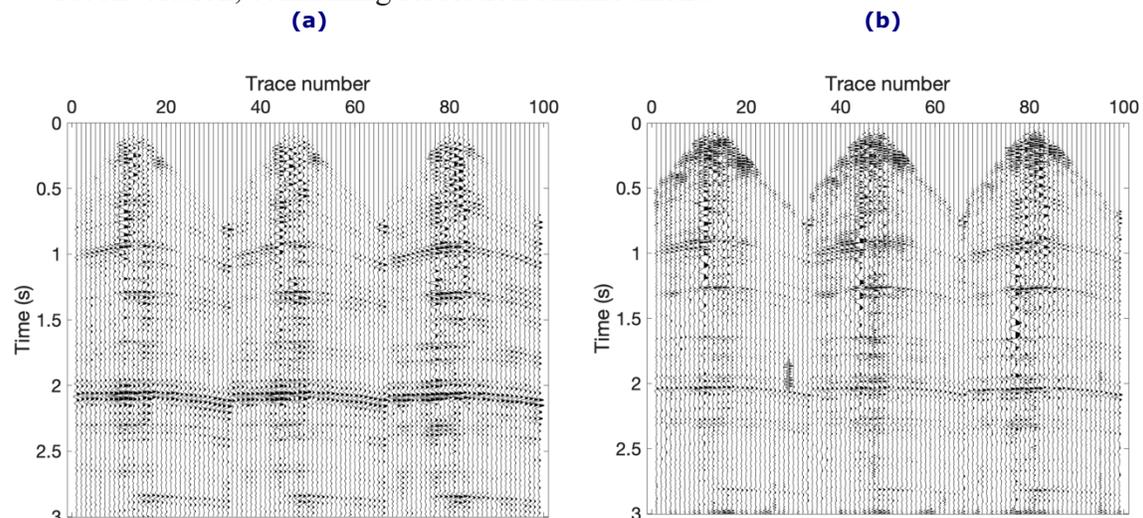

**Figure 4:** Comparison of shot gathers (a) before and (b) after applying spiking deconvolution using the Python code generated by ChatGPT for a 2D land-line from East Texas, USA. The prompt provided to GPT-4 was: '*Write a Python function based on Weiner optimum filtering to perform spiking deconvolution for a series of seismic shot records*.' The inputs to the function are seismic shot gather, length of the spiking filter, and pre-whitening percentage. The output of the function is the deconvolved seismic data.

2. *Innovative Idea Generation:* ChatGPT can inspire innovative thinking and generate creative ideas for research projects, fostering intellectual curiosity and exploration in various fields of study. I have already used it for inspiration to come up with new research questions that I further develop by reading the relevant literature.
3. *Improved Collaboration:* ChatGPT facilitates communication and collaboration among researchers and educators, enabling them to exchange ideas, share resources, and collaborate on projects or grant proposals more effectively. ChatGPT can quickly provide relevant information and insights on a wide range of topics. This helps researchers and educators from different disciplines or areas of expertise to better understand each other's work and contribute effectively to collaborative projects. Moreover, ChatGPT can provide concise explanations of complex concepts, making it easier for collaborators with diverse backgrounds to comprehend and engage in meaningful discussions. This enhances mutual understanding and facilitates effective teamwork.

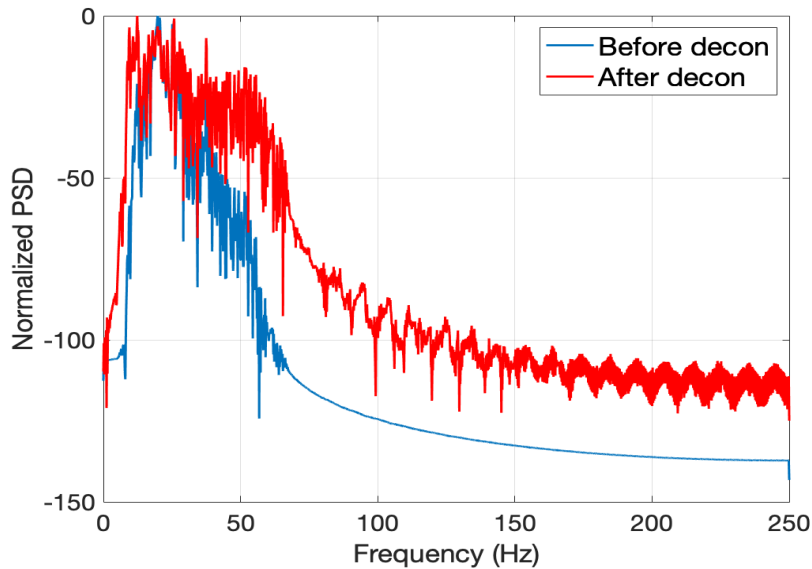

**Figure 5:** Normalized power spectral density of the average trace before (blue) and after (red) spiking deconvolution, confirming resolution enhancement.

4. *Curriculum Development:* ChatGPT can support educators in designing course materials, generating relevant examples, and crafting engaging activities or discussion questions, promoting a more dynamic and interactive learning environment. I have already used to it fine tune a course proposal on machine learning for geoscientists and engineers that I have recently been working on.
5. *Time-saving Administrative Assistance:* ChatGPT can help educators and researchers automate routine tasks, such as generate email responses, scheduling, or document organization, allowing them to dedicate more time to their core responsibilities.

*Challenges*: Despite the many benefits, there are numerous challenges associated with its use that researchers and educators should be cognizant of to fully harness its potential. Based on my personal experience, the key challenges include:
1. *Reliability:* While ChatGPT can efficiently process and synthesize vast amounts of data, it may occasionally provide outdated, incomplete, or inaccurate information. This necessitates users to verify and cross-check the information they receive with other credible sources, adding an extra layer of scrutiny to the research or teaching process.
2. *Ethical concerns:* Researchers and educators must be aware of the ethical implications associated with AI technologies, such as biased algorithms, invasive data collection practices, or the generation of misleading information. Addressing these concerns and ensuring responsible use is crucial in maintaining user trust and upholding the integrity of research and education.

3. *Potential for overreliance:* The convenience and efficiency provided by ChatGPT may lead to an overreliance on the AI for answers, potentially diminishing critical thinking and problem-solving skills among users. Researchers and educators must strike a balance between utilizing ChatGPT's capabilities and encouraging human reasoning, intellectual curiosity, and independent thought. This is particularly a serious challenge for educators to ensure their students do not fall into this trap.

*Lessons learned from writing a review paper using ChatGPT*: Recently, I tested ChatGPT's capabilities to write a review article on full waveform inversion (FWI; Waheed, 2023). Subsequently, the generated text was circulated among five experts in the field of FWI and their feedback was sought on the quality and correctness of information presented in the article. The experts unanimously agreed that the text is of high-quality and covered the topic quite comprehensively. It also provided some insights on the topic that were only obvious to those deeply familiar with the topic. Some of the important lessons learned through this activity are outlined below:

1. *Nonsensical references:* In most instances, ChatGPT generates nonsensical references that do not exist. This may happen as ChatGPT's primary goal is to generate human-like text that is coherent and contextually relevant. However, it may not always understand the specific implications of generating a citation. It generalizes from the patterns observed in its training data, which can result in the creation of citations that do not accurately reflect real sources. Moreover, while ChatGPT has an impressive capability to understand and process natural language, it does not possess a deep understanding of the content it generates. This means that it may not be able to distinguish between accurate and nonsensical citations, leading to the inclusion of the latter in its output.
2. *Erroneous information:* All the details and insights generated by ChatGPT were found to be correct, barring one slight error or inaccuracy. In FWI, the product of the forward-propagated wavefield and the back-propagated residuals in the adjoint method yields the gradient, not the sensitivity as it was claimed by ChatGPT. Nevertheless, the concepts are closely connected and, therefore, it is possible that ChatGPT picked this information up through a source that was misleading. This highlights the crucial role of expert supervision when reviewing and refining text generated by ChatGPT to ensure the accuracy and reliability of the final output.
3. *Generic and vague vocabulary:* In some sections of the article, ChatGPT used quite generic and vague vocabulary, which showed a lack of clarity or understanding of the topic. This may happen as ChatGPT is trained on vast amounts of text from various sources, including generic and non-specialized content. As a result, it may sometimes produce generic or vague vocabulary as it mimics the language patterns it has encountered. Moreover, while ChatGPT can process and understand natural language to a significant extent, its understanding of context may not always be as deep or nuanced as that of a human expert. This may quickly lead it to generated texts being contextually relevant but lacking specificity. Of course, ChatGPT is designed to generate text that is comprehensible to a wide range of users. Using generic or vague vocabulary allows it to maintain a balance between providing information and ensuring the content is accessible to users with varying levels of expertise in the subject matter.

Users should take the following precautions when using ChatGPT for writing tasks in general:
1. Provide clear and specific instructions or prompts to guide ChatGPT in generating more precise and specialized vocabulary.
2. Carefully review the AI-generated text and edit or revise it to replace generic or vague vocabulary with more precise and subject-specific terms.
3. Combine ChatGPT's output with expert knowledge to ensure that the writing is both accurate and specific to the topic at hand.

### III. Graduate Student (KA)

*Overall assessment*: As a Geophysics graduate student focusing on Computational Engineering for my master's degree, I find ChatGPT to be an extremely helpful resource for improving my coding skills, summarizing technical information, and gaining inspiration for small projects and writing tips. ChatGPT offers valuable guidance on coding practices, helps me quickly understand new concepts, and provides ideas for best research projects. Overall, leveraging ChatGPT and other emerging large language models (LLMs) can help one enhance creative thinking, sharpen coding proficiency, and accomplish research objectives with higher quality and productivity. But in spite of being a helpful resource, it is vital to remember that ChatGPT (and other language models) are just generative algorithms, and their output may sometimes be inaccurate or unreliable. Therefore, the outputs must always be assessed and questioned.

*Benefits*:
1. *Faster problem-solving:* ChatGPT can help quickly identify and resolve coding errors, saving time and frustration in the development process. Its ability to generate human-like text can make it feel like the user is collaborating with an experienced programmer, helping find solutions to problems more efficiently.
2. *Increased productivity:* ChatGPT can suggest ways to optimize the workflow and improve coding practices, which can help individuals become productive programmers. By identifying ways to streamline the work and reduce unnecessary steps, users can focus on writing high-quality code that meets the project goals.
3. *Personalized advice:* ChatGPT can be customized to your specific needs, allowing you to receive advice tailored to your project and goals. This personalized approach can help you learn faster and more effectively in your programming work.
4. *Unbeatable accessibility:* The availability of ChatGPT makes it a convenient and accessible resource for learning and problem-solving. Moreover, it always provides users with guidance and support whenever required, including off-hours, such as in the middle of the night or on weekends. In that regard, collaboration with ChatGPT is easier than collaborating with humans.

*Challenges*: While ChatGPT can be a valuable resource for improving coding skills and providing insights, there are potential threats associated with its use. One major issue is that ChatGPT can sometimes generate false information with high confidence, which can be misleading and result in errors or wasted time. Additionally, ChatGPT's limited dataset may only sometimes be up-to-date or reflect the latest advancements in specific fields. This means that relying solely on ChatGPT for guidance may not be the best approach for keeping up with rapidly changing technologies. Therefore, it is essential to be aware of these potential threats when using ChatGPT and to carefully evaluate its output before making decisions based on its recommendations. Without a deep understanding of the subject matter, evaluating the accuracy and relevance of ChatGPT's output can be challenging. Therefore, it is recommended to use ChatGPT as a supplement rather than a replacement for daily use and research purposes.

*Personal use case of ChatGPT:* As a master's student, I use ChatGPT to enhance the quality and efficiency of my research workflow in several ways, including:
1. *Summarizing information:* ChatGPT is a great tool to summarize information, such as articles, research proposals, etc. This helps me digest the information quickly and conclude on the specific topic. As a graduate student who has a hectic schedule, with this, I can wisely manage my time and focus on the most relevant and essential points from various sources of literature.
2. *Inspiration for implementation:* As they are generative algorithms, one of the best applications of the large language models (LLMs) is to get inspiration. This gives me inspiration for employing various programming languages, frameworks, or tools to carry out my study endeavor. ChatGPT can generate code snippets or pseudocode based on natural language descriptions or examples that I provide. ChatGPT can also suggest alternative or optimal coding methods that can improve my code's performance or readability.
3. *Learning specific modules or frameworks:* In the era of information abundance, finding the most suitable and reliable source of knowledge for learning specific topics can pose a significant challenge for researchers. This is especially true when they must develop scripts for their research projects involving complex or novel concepts and techniques. Therefore, there is a need for effective and efficient methods and tools that can assist researchers in acquiring the relevant skills and knowledge for their research domains. ChatGPT can generate well-documented code with comments or docstrings explaining the logic and functionality of each line or block. ChatGPT can also generate code that follows best practices and standards for different programming languages.

**Discussion**

We asked ChatGPT about its capabilities. Here are the unaltered responses to the prompts (**in bold print**) from our Q&A session:

**What are you good at?** As an AI language model, I'm designed to be proficient in several areas, such as: Natural Language Processing (NLP): I can understand and interpret human language, including grammar, syntax, and semantics.
1. *Conversational AI*: I can engage in human-like conversations, answer questions, and provide assistance.

2. *Knowledge Management*: I can store and retrieve vast amounts of information from my database and answer questions related to various topics.
3. *Language Translation*: I can translate text from one language to another, facilitating communication between people who speak different languages.
4. *Text Analysis:* I can analyze text and extract insights such as sentiment, intent, and key topics.

Overall, I'm good at providing information and assistance in a wide range of areas, including education, healthcare, business, and more.

***Is that all?*** No, those are just a few examples of the tasks I'm good at. In addition to the ones I mentioned, I can also perform a variety of other language-related tasks, such as:
1. *Named Entity Recognition*: identifying and classifying entities in a text, such as people, organizations, and locations.
2. *Text classification:* categorizing text into predefined categories, such as spam classification or sentiment analysis.
3. *Chatbot development*: engaging in a conversation with a user and responding appropriately.
4. *Speech recognition*: transcribing spoken language into text.
5. *Language modeling*: generating coherent text in a specific style or genre.
6. *Text completion*: generating plausible text based on a given prompt.

These are just a few examples of the many language-related tasks I can perform, but my abilities continue to evolve and expand as new techniques and technologies are developed.

***And?*** In addition to language-related tasks, I can also perform other types of tasks such as:
1. *Image recognition*: identifying objects, people, and other details in an image.
2. *Recommendation systems*: providing recommendations to users based on their preferences and behavior.
3. *Predictive modeling*: predicting future outcomes based on historical data.
4. *Time series analysis*: analyzing data over time and making predictions based on trends and patterns.
5. *Data mining*: discovering patterns and insights from large datasets.

These are just a few examples of the wide range of tasks that I'm capable of performing. However, my abilities are limited to what I've been trained on and the quality of the data I've been provided, so I may not be able to perform every task with the same level of accuracy and efficiency as a human expert in that particular domain.

***Can you compute?*** Yes, I can perform computations, including mathematical calculations, logical operations, and other types of numerical analysis. I can also work with various types of data, including text, images, and numerical data, and perform computations on that data. However, it's important to note that my abilities in computation are limited to what I've been programmed to do, and I may not be as efficient or accurate as specialized tools or software designed specifically for complex computational tasks. Nonetheless, for many common types of calculations and data analysis tasks, I can be a useful tool.

## Conclusions

September 2021 is the cut-off date for ChatGPT data processing, therefore, it has no knowledge of an ongoing war in Ukraine, or anything published scientifically after the cut-off date. Also, when given increasingly complexly formulated instructions that are logically correct, the response will increasingly falter, fail or are plainly wrong.

Overall, if you let ChatGPT work on text you give it as a starting point, it will track change. However, when conducting a deep-learning search of outside information, it will concoct and return text that is based on unknown correlations and filters. For example, when asked to write a biosketch of one of us, it created a myth that the person in question was from another nationality than he was, had been CEO of a major company and sold his company stake in 2014, among many other false claims. Care should be taken when using the results of any deep-search performed by ChatGPT – chances are high that some of its assertions will be blatantly wrong.

**Acknowledgement -** The authors acknowledge the use of ChatGPT, a large language model developed by OpenAI, in preparing this article, and the generous support provided by the College of Petroleum Engineering & Geosciences (CPG).